\documentclass[aps,prb,twocolumn,superscriptaddress,showpacs]{revtex4}

\usepackage{graphicx}
\usepackage{dcolumn} 
\usepackage{bm}      

\newcommand{\Tc}{{T$_c$~}}

\newcommand{\Hc}{{H$_c$$_2$ }}

\begin{document}

\preprint{APS/123-QED}

\title{Quasi-2D superconductivity and Fermi-liquid behavior in bulk CaC$_6$}

\author{E. Jobiliong}

\author{H.~D.~Zhou}

\author{J.~A.~Janik}

\author{Y.-J.~Jo}

\author{L.~Balicas}

\author{J.~S.~Brooks}

\author{C.~R.~Wiebe}
\email{cwiebe@magnet.fsu.edu} \affiliation{Department of Physics,
Florida State University, Tallahassee, FL 32306-3016, USA}
\affiliation{National High Magnetic Field Laboratory, Florida State
University, Tallahassee, FL 32306-4005, USA}

\date{\today}

\begin{abstract}

The intercalated graphite superconductor CaC$_6$ with \Tc $\sim$
11.5 K has been synthesized and characterized with magnetoresistance
measurements. Above the transition, the inter-plane resistivity
follows a T$^2$ dependence up to 50 K, indicative of Fermi liquid
behavior. Above 50 K, the data can be fit to the Bloch-Gr\"{u}neisen
model providing a Debye temperature of $\theta_D$ = 263 K. By using
McMillan formula, we estimate the electron-phonon coupling constant
to be $\lambda$= 0.85 which places this material in the
intermediate-coupling regime.  The angular dependence of the upper
critical field parallel and perpendicular to the superconducting
planes suggests that this material is a quasi-2D superconductor. All
of these measurements are consistent with BCS-like
superconductivity.

\end{abstract}

\maketitle

Despite the observation of superconductivity in graphite
intercalation compounds (GICs) over four decades ago, little
progress has been made to significantly raise the transition
temperature \Tc in this class of
materials.\cite{Hennig59,Hannay65,Koike78,Kobayashi81}  Current
theories support a model in which \Tc increases with increased
charge transfer from the intercalant to the graphene layers.
However, some members of this series seem to contradict this view.
For example, in LiC$_6$, the charge transfer is larger than that of
KC$_8$, but there is no evidence of superconductivity in
LiC$_6$.\cite{Rabii89} The application of high pressure also has
been used to raise the critical temperature, although in many cases
opposite phenomena is observed, except in the case of KC$_8$ which
successfully increases the critical temperature up to 1.5 K in 13
kbar (\Tc= 0.15 K in ambient pressure). From a theoretical approach,
superconductivity in GIC is an interesting problem since the
constituent elements are not superconducting alone. Recent works
have attacked this problem from the view of band structure
calculations.\cite{Jishi83,Ohno83,Csanyi05} It is the general
consensus of the community that finely tuned electron-phonon
interactions gives rise to BCS-like superconductivity in GICs, which
limits the maximum value of \Tc. However, recent interest in
low-dimensional materials such as MgB$_2$, a BCS superconductor with
a very high transition temperature \Tc = 39 K,\cite{Nagamatsu01} has
reinvigorated this field and the search for new materials with
finely tuned properties.  The discovery of relatively high \Tc s in
materials such as YbC$_6$ and CaC$_6$ (8 K and 11.5 K
respectively)\cite{Weller05,Emery05} provides an even further
impetus for the understanding of BCS-like phenomena in low
dimensional structures.

This paper details the extensive upper critical field
magnetoresistivity measurements on the new GIC superconductor
CaC$_6$. The results of these experiments include: (1)  The \Hc
values are determined for applied magnetic fields parallel and
perpendicular to the graphite planes. (2)  The resistivity as a
function of temperature is fit to several different models.
Fermi-liquid behavior is noted below 50 K (T$^2$ dependence), and
above 50 K, the best fit to the data is with a Bloch-Gr\"{u}neisen
model. The extracted coupling constant $\lambda$ agrees with density
functional theory calculations. (3) Through the dependence of the
upper critical field as a function of field direction, we have
determined that CaC$_6$ is a quasi-2D superconductor.

CaC$_6$ samples were prepared using highly-oriented pyrolytic
graphite with a liquid-solid reaction extreme.{\cite{Emery05}  A
lithium-calcium alloy (of the ratio 3:1) was prepared inside of an
argon glove box at 220 degrees C, and thin sheets of pristine
graphite were inserted.  The entire sample mixture was sealed in a
stainless-steel reaction container, and then placed on a hot plate
at 350 degrees for 10 days.  The final samples were extracted from
the molten solution inside of the glove box, and only very thin
samples which exhibited shiny metallic surfaces were used for the
measurements. The transition temperature of 11.5 K was confirmed
with DC susceptibility measurements (Quantum Design SQUID) in a
field of 50 G applied parallel to the ab-plane. From the saturation
of the diamagnetic signal, the samples used for the remaining
measurements were estimated to have a volume fraction of over 95
percent of the superconducting phase.  The resistivity data were
measured using a conventional four-probe method with current applied
along c-axis. The size of a single crystal is 1.5 mm$\times$1
mm$\times$0.2 mm.The resistivity measurements were completed in a
He-flow cryostat at the NHMFL, Tallahassee.

A few theoretical models have been developed to explain the
temperature dependence of the inter-plane resistivity in GICs. One
of these is inspired by the theory of variable-range hoping in
parallel with band conduction.\cite{Powers88} This model can be well
applied for acceptor GICs. Another model proposed by
Sugihara\cite{Sugihara84} suggests the importance of impurity and
phonon-assisted hoping that has a linear dependence on temperature.
We estimate the conductivity (along c-axis) in this sample to be
$8.7\times10^{3}~ \Omega^{-1}\mathrm{cm}^{-1}$, which is a typical
value for donor GICs. Thus, we use the latter model in addition with
the theory of thermal scattering of charge carriers in a single
band, which has been used for many GICs\cite{Rosenberg63} to analyze
the temperature dependence of the inter-plane resisitivity above 50
K, as shown in figure \ref{resist}. According to this model the
total resistivity can be written as,
\begin{equation}\label{rho}
\rho(T)=\rho_0+ AT + BT^5\displaystyle\int^
{\theta_D/T}_0\frac{x^5}{(e^x-1)(1-e^{-x})}\, dx
\end{equation}
where $\theta_D$ is the Debye temperature. The first term in
Eq.\ref{rho} is a temperature-independent constant, the second term
is related to the phonon-assisted hoping and the last term is
related to the electron-phonon scattering, also known as the
Bloch-Gr\"{u}neisen formula. This parameter from our fit is found to
be 263(1) K, which is in good agreement with other GICs fit with
this model (which range from 200 K - 300
K).\cite{Koike78,Kobayashi85,Pendrys81} TThe inset shows the
normalized resistivity at lower temperatures and as a function of
T$^2$. The clear dependence on T$^2$ is strong evidence for
Fermi-liquid behavior for CaC$_6$.

Using this characteristic temperature, one can estimate the
electron-phonon coupling parameter using the
McMillan\cite{McMillan68} equation:
\begin{equation}
\lambda=\frac{\mu\ln(\frac{1.45T_{C}}{\theta_{D}})-1.04}{1.04+\ln(\frac{1.45T_{C}}{\theta_{D}})(1-0.62\mu)}
\end{equation}
where $\mu$ is the screened potential and $\lambda$ is the
electron-phonon coupling parameter.  Using the $\mu$ parameter of
0.1, a value of $\lambda$ of 0.85 is obtained.  This is in excellent
agreement with theoretical predictions using density functional
theory (0.83).\cite{Calandra05} The high value of $\lambda$
indicates that this material is in the intermediate-coupling regime.
Other superconducting GICs typically have much lower values for
$\lambda$ = 0.2 - 0.5.\cite{Koike78,Alexander81,Iye82}  The high
value of $\lambda$
 certainly plays a role in the anomalously high \Tc compared to
 other intercalated graphite compounds.
\begin{figure}[tbp]
    \centering
    \includegraphics[scale=.45,angle=270]{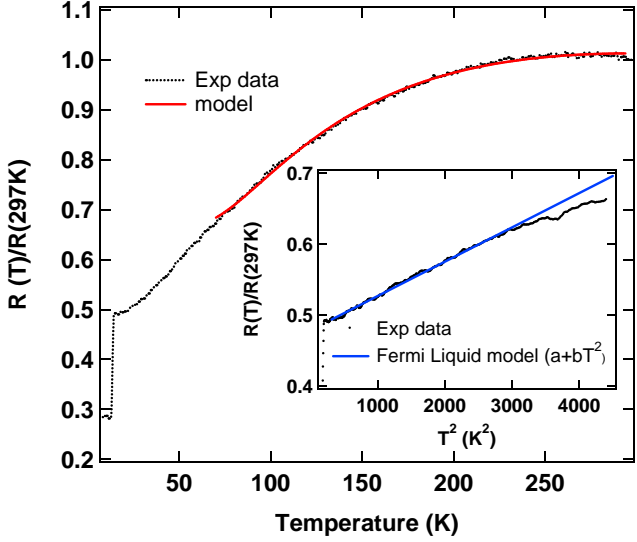}
    \caption{Inter-plane resistivity as a function of temperature normalized to 297
K.  The inset shows a clear T$^2$ dependance below 50 K.  Above 50
K, the data is fit to the Bloch-Gr\"{u}neisen model as outlined in
the text.} \label{resist}
\end{figure}

 The inter-plane resistivity measurements as a function of applied field are
shown in the inset of figure \ref{fielddep}.  The midpoint of the
gradient of the resistivity drop was used to estimate the transition
temperature \Hc perpendicular and parallel to the ab-plane.  The
values of \Hc agree with susceptibility measurements for CaC$_6$ in
the previous work.\cite{Emery05} As seen in other GIC
compounds,\cite{Iye82,Iye82S,Chaiken90} a linear dependence of \Hc
on temperature is observed, as shown in figure \ref{phase diagram}.
This linear dependence indicates that the anisotropy is temperature
independent. Moreover, this also suggests that there is no evidence
of dimensional cross-over, at least down to 1.4 K. The critical
field in the anisotropic Ginzburg-Landau theory can be written
as\cite{Lawrence71}
\begin{equation}\label{coh length}
H_{C2}^i = \frac{\Phi_{0}}{2\pi\xi_{j}(T)\xi_{k}(T)} =
\frac{\Phi_{0}}{2\pi\xi_{j}(0)\xi_{k}(0)}\Big(1-\frac{T}{T_{c}}\Big)
\end{equation}
where $\Phi_0=h/2e=2.07\times10^{-15}$ Tm$^{2}$ is the flux quantum
and $\xi$ is the coherence length. The indices $\textit{i}$,
$\textit{j}$ and $\textit{k}$ represent the cyclic permutation of
the directions $\textit{a}$, $\textit{b}$ and $\textit{c}$. By
applying Eq. \ref{coh length} for our data in figure \ref{phase
diagram}, we obtain the correlation lengths $\xi_{\perp}$(0) and
$\xi_{//}$(0) to be 6.0 nm and 29.7 nm, which can be compared to DC
susceptibility measurements of 13.0 nm and 35.0 nm.\cite{Emery05} It
was originally surmised that this was an anisotropic 3D
superconductor, but the c-axis correlation length has been
over-estimated in the previous work.\cite{Emery05} Since the c-axis
lattice parameter is 1.357 nm due to the stacking sequence of Ca
atoms, it is perhaps not unusual that this material may have
quasi-2D superconductivity.

\begin{figure}[t]
\linespread{1}
\par
\begin{center}
\includegraphics[scale=.45,angle=270]{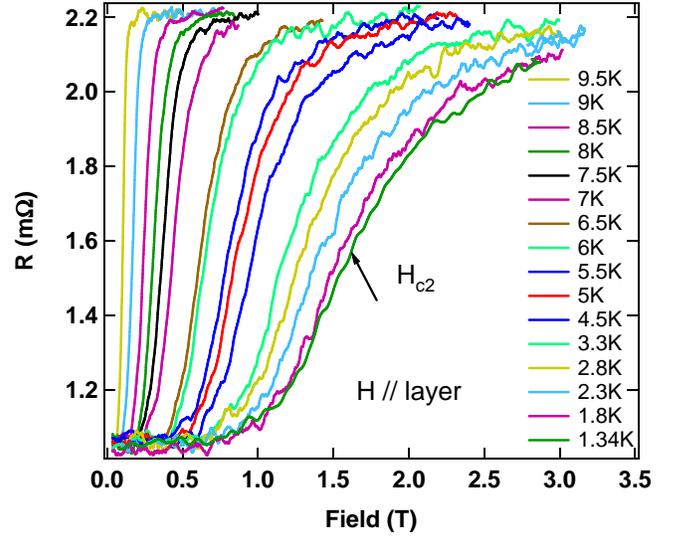}
\end{center}
\par
 \caption{The inter-plane resistivity plots as
    a function of applied field which were used to extract the critical
    field as the midpoint of the resistivity drop at the transition.}
    \label{fielddep}
\end{figure}

\begin{figure}[t]
\linespread{1}
\par
\begin{center}
\includegraphics[scale=.45,angle=270]{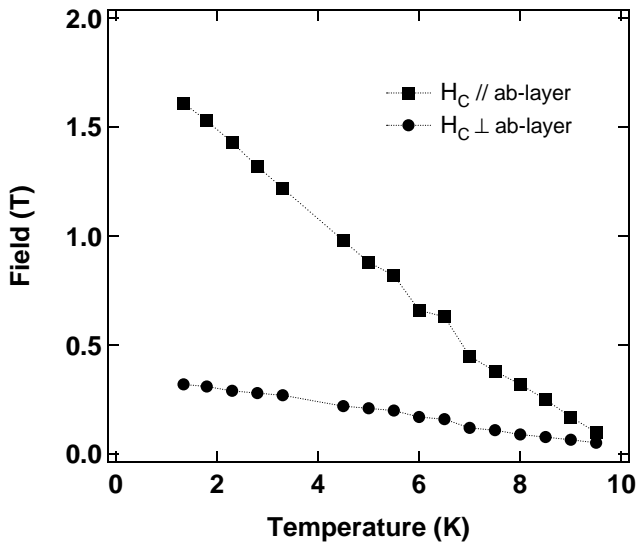}
\end{center}
\par
\caption{Upper critical field \Hc for fields perpendicular and
parallel to ab-plane as a function of temperature} \label{phase
diagram}
\end{figure}

The angular dependence of the critical field in CaC$_6$, which has a
higher critical field parallel to the ab-plane than that of the
perpendicular direction, is shown in Figure \ref{phase diagram}.
Note that the 90 degree position corresponds to an applied field
parallel to the ab-plane.  We compare this behavior with the two
following models.  The first model is using anisotropic
Ginzburg-Landau (GL) theory, which is valid when the interlayer
spacing is much smaller than the c-direction coherence length.  In
this case, the upper critical field depends on the angle between the
layers and the applied field through \cite{Morris72}
\begin{equation}\label{GL}
\Big[\frac{H_{c2}(\theta)cos(\theta)}{H_{c2\bot}}\Big]^2+\Big[\frac{H_{c2}(\theta)sin(\theta)}{H_{c2//}}\Big]^2
=1
\end{equation}
where the upper critical fields for directions parallel and
perpendicular to the ab-plane are H$_{c2//}$ and H$_{c2\bot}$. The
second model is based on Lawrence-Doniach (LD) theory, which assumes
that there is a weak coupling between the superconducting layers in
the two-dimensional (2D) limit.  In this case, the angular
dependence of the layers is found to be:\cite{Tinkham63}
\begin{equation}\label{LD}
\Big|\frac{H_{c2}(\theta)cos(\theta)}{H_{c2\bot}}\Big|+\Big[\frac{H_{c2}(\theta)sin(\theta)}{H_{c2//}}\Big]^2
=1
\end{equation}
This model has been used to describe the angular dependence of the
magnetoresistivity for thin films, and for two-dimensional
superconductors in general.\cite{Schneider93}  The main feature in
GL model is a rounded shape at 90 degrees, while the LD model
produces a sharp cusp at 90 degrees. We fit these two models to our
data, as shown in figure \ref{uppercrit2}. The Lawrence-Doniach
model gives a better fit than that of Ginzburg-Landau model.
Furthermore, a cusp-like feature at 90 degrees, which is a signature
of the LD-model, is also observed as shown in the inset of figure
\ref{uppercrit2}. We fit the data at low temperatures (1.4 K) using
these two models.  Note that the Lawrence-Doniach model still gives
a better fit up to 180 degrees, and the cusp-like is less
pronounced.  We confirmed this 2D-model for another measurement in a
helium-3 system at a temperature of 0.5 K (data is not shown).

\begin{figure}[t]
\linespread{1}
\par
\begin{center}
\includegraphics[scale=.45,angle=270]{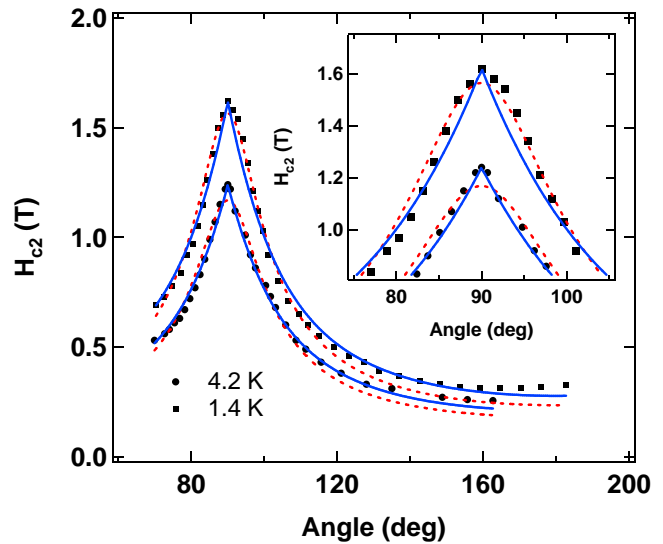}
\end{center}
\par
\caption{Angular dependence of the upper critical field for 1.4 K
and 4.2 K. The solid and dashed lines indicate the fitting curves
for LD-model and GL-model, respectively.} \label{uppercrit2}
\end{figure}

One theory that has been used to understand the mechanism of
superconductivity in GICs is that proposed by Jishi.\cite{Jishi83}
In this model, it has been presumed that the superconductivity
arises from a coupling between the graphene \textit{$\pi$} bands and
the intercalant layer \textit{s} band. Moreover, the linear
dependence of critical field also can be explained quantitatively
using this model.\cite{Jishi92} However, this model is valid only in
the weak-coupling regime ($\lambda < 0.4$) and our results indicate
that CaC$_6$ is in intermedite coupling regime ($\lambda < 0.85$).
Although other models have been proposed to explain the origin of
superconductivity in CaC$_6$
recently,\cite{Csanyi05,Calandra05,Mazin05} there is no quantitative
explanation for the linear dependence of the upper critical field.

The observation of a T$^2$ dependence of the resistivity is
consistent with the conjecture that CaC$_6$ is a BCS-like
superconductor. Recent penetration depth measurements\cite{Lamura06}
and heat capacity experiments\cite{Kim06} have showed that the
superconductivity is indeed s-wave and BCS-like, respectively.
Fermi-liquid theory provides a natural avenue to produce BCS
superconductivity.  It is surprising that such a large transition
temperature of 11.5 K is observed, but given the large coupling
parameter deduced from our measurements (and, recently through
specific heat experiments), it is likely that the origin of the
superconductivity is through finely tuned electron-phonon
interactions.

In conclusion, extensive resistivity measurements on the new GIC
CaC$_6$ have revealed several key features of this new
superconductor: (1) There is a prominent anisotropy in the upper
critical field parallel and perpendicular to the layers, (2) there
is a Fermi-liquid regime below 50 K and (3) this GIC is a quasi-2D
superconductor. BCS-theory gives a value for the coupling constant
which is in the intermediate regime, which is in agreement with
recent specific heat measurements.\cite{Kim06}  All of these
observations are consistent with the view that CaC$_6$ is a
BCS-superconductor with finely tuned electron-phonon interactions
which gives rise to the large \Tc.

\begin{acknowledgments}
This research was sponsored by the National Nuclear Security
Administration under the Stewardship Science Academic Alliances
program through DOE Research Grant \#DE-FG03-03NA00066, NSF Grant
\#DMR-0203532, NSF Grant \#DMR-0449569 and the NHMFL is supported by
a contractual agreement between the NSF and the State of Florida.
\end{acknowledgments}

\bibliography{Bibli_CaC6}

\end{document}